\documentclass{article}
\usepackage{spconf,amsmath,graphicx}
\usepackage{amsfonts}
\usepackage{amsmath,amssymb}
\usepackage{adjustbox}
\usepackage{algorithmicx}
\usepackage{algpseudocode}
\usepackage[ruled]{algorithm}
\usepackage{graphicx}
\usepackage{caption}
\usepackage{graphicx,times,amsmath} 
\usepackage{caption}
\usepackage{subcaption}
\usepackage[rightcaption]{sidecap}
\usepackage{caption}
\usepackage{multirow}
\usepackage{hhline}
\usepackage{amsmath}
\usepackage{epstopdf}
\usepackage{tabularx}
\usepackage{tikz}
\usetikzlibrary{matrix}
\usepackage{mathtools, nccmath}
\usepackage{pst-node}
\graphicspath{ {images/} }

\usepackage{alphalph}

\title{Deep Sequential Learning for Cervical Spine Fracture Detection on Computed Tomography Imaging}
\usepackage{fancyhdr}
\begin{document}
\newcommand*{\img}{%
  \includegraphics[
    width=\linewidth,
    height=20pt,
    keepaspectratio=false,
  ]{example-image-a}%
}

\name{\begin{tabular}{c}Hojjat Salehinejad$^{1,2,3,*}$, Edward Ho$^{4}$, Hui-Ming Lin$^{4}$, \textit{Priscila Crivellaro}$^{4}$, \textit{Oleksandra Samorodova}$^{4}$, \\
\textit{Monica Tafur Arciniegas}$^{4}$, \textit{Zamir Merali}$^{5}$, \textit{Suradech Suthiphosuwan}$^{4}$, \textit{Aditya Bharatha}$^{4}$, \\
\textit{Kristen Yeom}$^{6}$, \textit{Muhammad Mamdani}$^{1,2,7}$, \textit{Jefferson Wilson}$^{5}$, and \textit{Errol Colak}$^{1,4}$\end{tabular}
\vspace{-4mm}
}
\address{
$^{1}$LKS-CHART, St. Michael’s Hospital, Toronto, Canada\\
$^{2}$St. Michael's Hospital, Unity Health Toronto, Toronto, Canada\\
$^{3}$Department of Electrical \& Computer Engineering, University of Toronto, Toronto, Canada\\
$^{4}$Department of Medical Imaging, University of Toronto, Toronto, Canada\\
$^{5}$Division of Neurosurgery, Department of Surgery, University of Toronto, Toronto, Canada\\
$^{6}$Department of Radiology, Stanford University, Stanford, California, USA\\
$^{7}$Temerty Faculty of Medicine, University of Toronto, Toronto, Canada\\
\textit{$^{*}$Email: hojjat.salehinejad@unityhealth.to}
\vspace{-4mm}}

\maketitle
%

\begin{abstract}
Fractures of the cervical spine are a medical emergency and may lead to permanent paralysis and even death. Accurate diagnosis in patients with suspected fractures by computed tomography (CT) is critical to patient management. In this paper, we propose a deep convolutional neural network (DCNN) with a bidirectional long-short term memory (BLSTM) layer for the automated detection of cervical spine fractures in CT axial images. We used an annotated dataset of 3,666 CT scans (729 positive and 2,937 negative cases) to train and validate the model. The validation results show a classification accuracy of $70.92\%$ and $79.18\%$ on the balanced (104 positive and 104 negative cases) and imbalanced (104 positive and 419 negative cases) test datasets, respectively.
\end{abstract}
\begin{keywords}
Cervical spine, deep learning, fracture detection.
\end{keywords}
\section{Introduction}
\label{sec:intro}
The cervical spine consists of seven stacked bones called vertebrae, labeled C1 through C7, intervertebral discs, and ligaments. The top of the cervical spine articulates with the skull, and the bottom connects to the thoracic spine. Trauma to the cervical spine results in over 1 million emergency department visits per year in North America~\cite{dunsker2019deep}. Severe trauma from falls, motor vehicle accidents, and sports injuries can result in fractures and/or dislocations of the cervical spine~\cite{bland1990anatomy}. Cervical spine fractures may lead to permanent paralysis from spinal cord injury or even death~\cite{bland1990anatomy}. A computed tomography (CT) scan of the cervical spine is the most common and preferred imaging modality used to detect fractures. 

Despite advances in machine learning, very limited methods have been proposed for fracture detection in cervical spine CT images. A 3D ResNet-101~\cite{he2016deep} deep convolutional neural network (DCNN) was utilized in~\cite{dunsker2019deep}, which was trained on 990 normal and 222 fracture cases. Performance of this method with respect to Area Under the Receiver Operating Characteristic (AUROC) and Area Under Precision-Recall Curve (AUPRC) metrics on the validation dataset (98 normal and 37 with fracture cases) at image level was AUROC=0.87 and AUPRC=0.52 and at case level was AUROC=0.85 and AUPRC=0.82~\cite{dunsker2019deep}.

Automated fracture detection in cervical spine images is a very challenging problem. Cervical spine CT scans with a fracture are highly imbalanced in terms of normal images and those with a fracture. Different methods such as spatial transformations~\cite{salehinejad2018image}, spatial-temporal transformations~\cite{salehinejad2018cylindrical}, and generative adversarial models~\cite{salehinejad2018synthesizing} can be used to reduce this imbalance. In addition, publicly available annotated cervical spine data is very limited, making advances to automated cervical spine fracture detection even more difficult. To address this, our team has annotated a dataset of cervical spine CT images (see Subsection~\ref{sec:data}) and we are in process of releasing this dataset to the research community\footnote{For updates visit dila.ai.}.   
In this paper, we model fracture detection as a classification problem and propose a DCNN with bidirectional long short-term memory (BLSTM) layer as a baseline model to address this problem on axial cervical spine CT images.

\begin{figure*}[!th]
\centering
\captionsetup{font=footnotesize}
\includegraphics[width=0.85\textwidth]{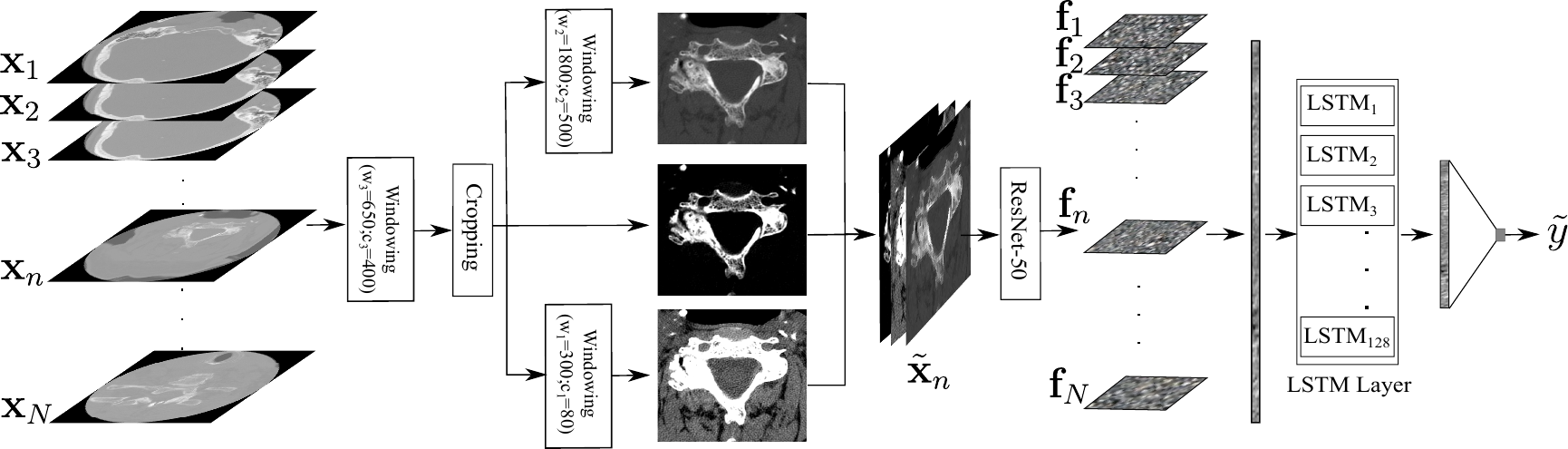}
\caption{Proposed model for fracture detection on cervical spine scans. The input scan has $N$ axial images and the BLSTM layer has 128 LSTM units.}
\label{fig:convlstm_model}
\vspace{-4mm}
\end{figure*}

\section{Fracture Detection in Cervical Spine}
\label{sec:proposedmethod}
 A cervical spine CT has $N$ number of axial image slices along the cranio-caudal axis where we represent each image with a vector $\mathbf{x}_{n}$. Therefore, we can model the cervical spine as the set of input images $\mathbf{X}=(\mathbf{x}_{1},\mathbf{x}_{2},...,\mathbf{x}_{N})$ with the corresponding image level labels $\mathbf{y}=(y_{1},y_{2},...,y_{N})$, where $y_{n}=1$ means the image contains at least one fracture and $y_{n}=0$ is the opposite. We can also define a case level label $y\in\{0,1\}$, where if $y=1$ at least one image contains a fracture and if $y=0$ none of the images contain a fracture. Figure~\ref{fig:convlstm_model} shows different steps of the proposed model. It has two major steps, which are prepossessing of the input images $\mathbf{X}$ and learning a mapping function from the preprocessed images to the target $y$. Different steps are discussed as follows.

\begin{figure}[!t]
\centering
\captionsetup{font=footnotesize}
\includegraphics[width=0.4\textwidth]{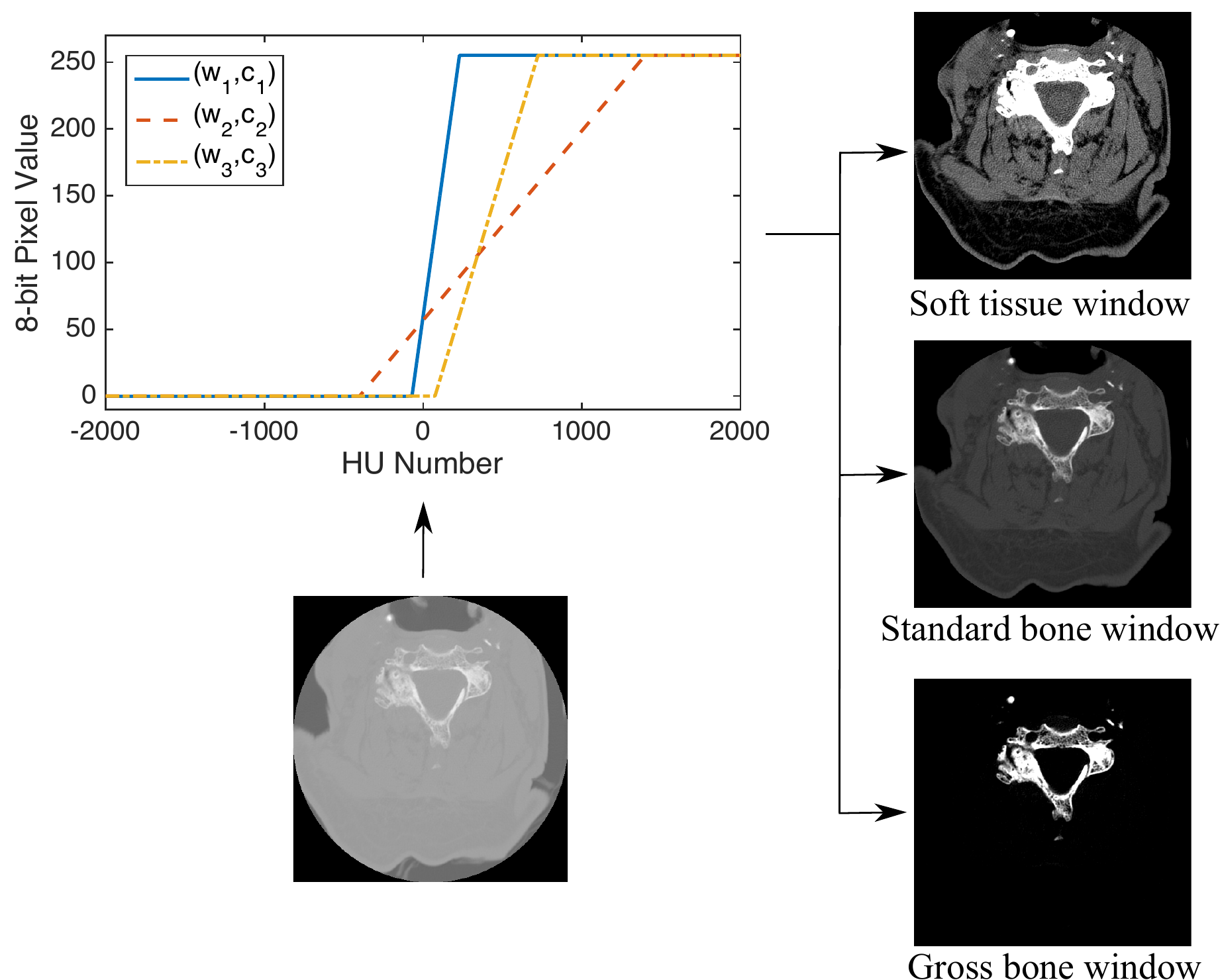}
\caption{Windowing of a given cervical spine CT image (in Hounsfield units (HU)) according to the following rules: 1) Soft tissue window ($w_{1}=300;c_{1}=80$); 2) Standard bone window ($w_{2}=1,800;c_{2}=500$); 3) Gross bone window ($w_{3}=650;c_{3}=400$).}
\vspace{-5mm}
\label{fig:cspine_windowing}
\end{figure}

\subsection{Preprocessing}
 Each pixel $x_{i}\in\mathbf{x}_{n}$ is a quantitative representation of the radiodensity of different substances in the scanning area where $-1,000\lesssim x_{i}\lesssim 3,000$,~\cite{kalra2018developing}. 
 Fractures of the cervical spine are heterogeneous in their location, type, and composition. They are defined as a break in at least one bone in the cervical spine. The cervical spine can be malaligned as a result of a fracture and/or a ligamentous injury. Both fractures and ligaments injuries can be associated with localized bleeding and soft tissue swelling. In consultation with a panel of radiologists, three window width and center schemes are chosen to enhance cervical spine bone and surrounding tissues, which are soft tissue window $(w_{1}=300;c_{1}=80)$, standard bone window $(w_{2}=1,800;c_{2}=500)$, and gross bone window $(w_{3}=650;c_{3}=400)$. Figure~\ref{fig:cspine_windowing} shows the corresponding functions to the proposed windowing schemes to map $x_{i}$ to three different values $\tilde{x}_{i}^{(1)}$, $\tilde{x}_{i}^{(2)}$, $\tilde{x}_{i}^{(3)}$.
 The represented image with the gross bone window  $\tilde{\mathbf{x}}_{n}^{(3)}$ is then used to crop the images with $5\%$ margin from each side of detected cervical spine using Otsu's method~\cite{bangare2015reviewing}. The cropped images are normalized and resized (padded with zero if not square) to $384\times 384$.

\subsection{Learning and Inference}
For the image $\mathbf{x}_{n}\in\mathbf{X}$, the preprocessed images are concatenated as ${\tilde{\mathbf{x}}_{n}=(\tilde{\mathbf{x}}_{n}^{(1)}\oplus \tilde{\mathbf{x}}_{n}^{(2)}\oplus \tilde{\mathbf{x}}_{n}^{(3)})}$ where $\oplus$ is the concatenation operator. The cervical spine images have spatiotemporal dependency. First, ResNet-50~\cite{szegedy2016inception} is trained using randomly selected batches of images from all training cases. The objective is to train ResNet-50 as the function $\mathbf{F}=\phi(\tilde{\mathbf{X}})$ where $\mathbf{f}_{n}\in\mathbf{F}$ is the extracted feature map from  $\tilde{\mathbf{X}}_{n}\in\mathbf{X}$. As Figure~\ref{fig:convlstm_model} shows, the feature maps are vectorized and concatenated as $\tilde{\mathbf{f}}=(\mathbf{f}_{1}\oplus\mathbf{f}_{2}\oplus,...,\mathbf{f}_{N})$. The second phase is learning the temporal dependency among axial images using a bidirectional network of LSTM units to map $\tilde{\mathbf{f}}$ to the case label $y$. The loss is calculated using binary cross-entropy function with respect to the target label $y$. 
In the inference mode, the feature extractor $\phi(\cdot)$ and BLSTM layer work together for a given case $\mathbf{X}$ to generate the inference $\tilde{y}$.



\begin{table}[]
\centering
\captionsetup{font=footnotesize}
\caption{Distribution of negative and positive cervical spine fracture cases based on the gender and age.}
\begin{adjustbox}{width=0.3\textwidth}
\begin{tabular}{lll}
\multicolumn{1}{l}{} & Positive    & Negative   \\ \hline
\multicolumn{1}{l}{Female} &    $6.74\%$         &     $28.20\%$       \\ \hline
\multicolumn{1}{l}{Male}   &     $13.15\%$        &    $51.91\%$        \\ \hline\hline
\multicolumn{1}{l}{Age}   &     $59.42\pm22.91$        &    $50.40\pm21.7$        \\ \hline
\end{tabular}%
\end{adjustbox}
\label{T:age_gender_count}
\vspace{-4mm}
\end{table}

\begin{figure*}[!ht]
\centering
\centering
\captionsetup{font=footnotesize}
\begin{subfigure}[t]{0.24\textwidth}
\captionsetup{font=footnotesize}
\centering
\includegraphics[width=.8\textwidth]{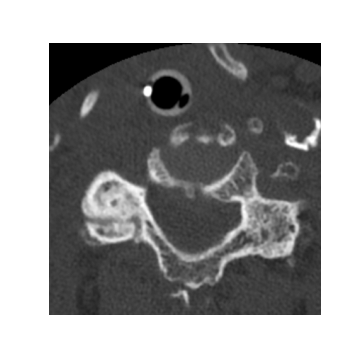}
\vspace{-4mm}
\caption{Axial image with three fractures.}
\label{fig:}
\end{subfigure}%
\begin{subfigure}[t]{0.24\textwidth}
\captionsetup{font=footnotesize}
\centering
\includegraphics[width=.8\textwidth]{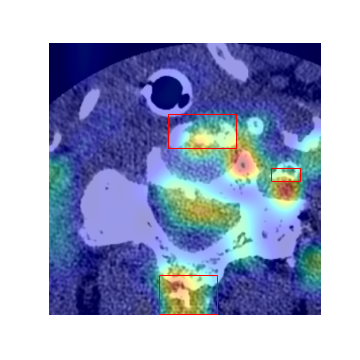}
\vspace{-4mm}
\caption{Heatmap of (a).}
\label{fig:}
\end{subfigure}%
~
\begin{subfigure}[t]{0.24\textwidth}
\captionsetup{font=footnotesize}
\centering
\includegraphics[width=.8\textwidth]{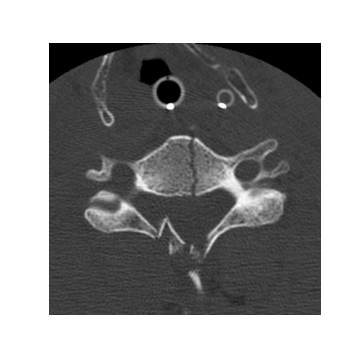}
\vspace{-4mm}
\caption{Axial image with two fractures.}
\label{fig:}
\end{subfigure}%
\begin{subfigure}[t]{0.24\textwidth}
\captionsetup{font=footnotesize}
\centering
\includegraphics[width=.8\textwidth]{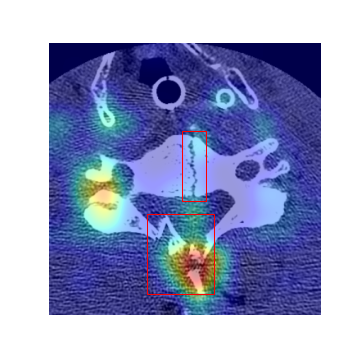}
\vspace{-4mm}
\caption{Heatmap of (c).}
\label{fig:}
\end{subfigure}%
\vspace{-2mm}
\caption{(a),(c): Samples of cervical spine axial images with fracture; (b),(d): Grad-CAM~\cite{selvaraju2017grad} heatmap from the last layer of ResNet-50 corresponding to (a) and (c), respectively. Ground-truth is represented with red boxes.}
\label{fig:heatmap_cspine}
\end{figure*}

\begin{table*}[!th]
\centering
\captionsetup{font=footnotesize}

\caption{Image level performance results on imbalanced (Imblcd.) (2,909 positive cases; 164,349 negative images; based on the distribution of training dataset) and balanced (Blcd.) (2,909 positive images; 2,909 negative images) test datasets with 7-fold cross-validation using ResNet-50. TPR: Sensitivity; TNR: Specificity; PPV: Positive predictive value; NPV: Negative predictive value; F1: F1 score; Acc: Accuracy; MCC: Matthews correlation coefficient; AUC: Area under the curve. All the values are in percent.}
\label{tab:my-table}
\begin{adjustbox}{width=.85\textwidth}
\begin{tabular}{|c|cccccccc|}
\hline
Data &
  TPR &
  TNR &
  PPV &
  NPV &
  F1 &
  Acc &
  MCC &
  AUC \\ \hline
\begin{tabular}[c]{@{}c@{}}Imblcd.\end{tabular} &
  $77.21\pm 4.1$ &
  $80.06\pm 3.0$ &
  $06.47\pm 0.6$ &
  $99.50\pm 0.1$ &
  $11.92\pm 1.0$ &
  $80.01\pm 2.9$ &
  $18.47\pm 1.0$ &
  $78.63\pm 1.2$ \\ \hline
\begin{tabular}[c]{@{}c@{}}Blcd.\end{tabular} &
  $77.21\pm 4.1$ &
  $77.62\pm 2.7$ &
  $13.78\pm 1.0$ &
  $98.67\pm 0.2$ &
  $23.35\pm 1.3$ &
  $77.61\pm 2.5$ &
  $26.11\pm 1.0$ &
  $77.42\pm 1.1$ \\  \hline
\end{tabular}%
\end{adjustbox}
\label{T:cspine_resnet50}
\end{table*}

\begin{table*}[!th]
\centering
\captionsetup{font=footnotesize}

\caption{Case level performance results on imbalanced (Imblcd.) (104 positive cases; 419 negative cases; based on the distribution of training dataset) and balanced (Blcd) (104 positive cases; 104 negative cases) test datasets with 7-fold cross-validation.  TPR: Sensitivity; TNR: Specificity; PPV: Positive predictive value; NPV: Negative predictive value; F1: F1 score; Acc: Accuracy; MCC: Matthews correlation coefficient; AUC: Area under the curve. All the values are in percent.}
\label{tab:my-table}
\begin{adjustbox}{width=1\textwidth}
\begin{tabular}{|c|c|cccccccc|}
\hline
Model &
  Data &
  TPR &
  TNR &
  PPV &
  NPV &
  F1 &
  Acc &
  MCC &
  AUC \\ \hline
\begin{tabular}[c]{@{}c@{}}ResNet-50 + BLSTM-96\end{tabular} &
  \multirow{2}{*}{\parbox[c]{2mm}{\multirow{2}{*}{\rotatebox[origin=c]{90}{Imblcd.}}}} &
  $64.19\pm 5.7$ &
  $78.67\pm 6.6$ &
  $43.62\pm 6.3$ &
  $89.83\pm 1.5$ &
  $51.66\pm 5.5$ &
  $75.79\pm 5.2$ &
  $37.84\pm 7.6$ &
  $71.43\pm 3.9$ \\ \cline{1-1} \cline{3-10} 
\begin{tabular}[c]{@{}c@{}}ResNet-50 + BLSTM-128\end{tabular} &
   &
  $62.28\pm 6.0$ &
  $80.84\pm 2.9$ &
  $44.83\pm4.8$ &
  $89.62\pm1.6$ &
  $52.06\pm4.9$ &
  $77.15\pm2.9$ &
  $38.54\pm6.7$ &
  $71.56\pm3.7$ \\\cline{1-1} \cline{3-10} 
  \begin{tabular}[c]{@{}c@{}}ResNet-50 + BLSTM-256\end{tabular} &
   &
  $59.01\pm 5.7$ &
  $84.12\pm 4.9$ &
  $48.54\pm6.7$ &
  $89.34\pm1.5$ &
  $52.92\pm4.6$ &
  $\mathbf{79.18\pm3.8}$ &
  $40.36\pm6.5$ &
  $71.57\pm3.1$ \\ \hline\hline
\begin{tabular}[c]{@{}c@{}}ResNet-50 +  BLSTM-96\end{tabular} &
  \multirow{3}{*}{\parbox[c]{2mm}{\multirow{1}{*}{\rotatebox[origin=c]{90}{Blcd.}}}} &
  $64.19\pm 5.7$ &
  $77.11\pm 7.3$ &
  $74.14\pm 5.4$ &
  $68.36\pm 3.1$ &
  $68.58\pm 3.8$ &
  $70.65\pm 3.5$ &
  $41.90\pm 7.1$ &
  $70.65\pm 3.5$ \\ \cline{1-1} \cline{3-10} 
\begin{tabular}[c]{@{}c@{}}ResNet-50 +  BLSTM-128\end{tabular} &
   &
  $62.28\pm 6.0$ &
  $79.84\pm 3.1$ &
  $75.55\pm 3.0$ &
  $68.06\pm 3.5$ &
  $68.17\pm 4.2$ &
  $\mathbf{71.06\pm 3.1}$ &
  $42.86\pm 5.9$ &
  $71.06\pm 3.1$ \\ \cline{1-1} \cline{3-10} 
\begin{tabular}[c]{@{}c@{}}ResNet-50 + BLSTM-256\end{tabular} &
   &
  $57.75\pm 4.9$ &
  $84.09\pm 5.3$ &
  $78.87\pm 4.9$ &
  $66.63\pm 1.8$ &
  $66.44\pm 2.8$ &
  $70.92\pm 1.9$ &
  $43.62\pm 4.3$ &
  $70.92\pm 1.9$ \\ \hline
\end{tabular}%
\end{adjustbox}
\label{T:cspine_resnet50lstm}
\end{table*}

\section{Results}
\label{sec:results}

\subsection{Data}
\label{sec:data}

The dataset contains cervical spine CT scans of 3,666 cases (729 positive and 2,937 negative cases) which corresponds to 1,174,335 images (20,392 fracture positive and 1,153,943 fracture negative). This dataset is pixel-level annotated by three expert radiologists and the intersection of annotated regions was reviewed by a neuroradiologist. Table~\ref{T:age_gender_count} shows the distribution of gender and age of cases over the fracture positive and negative classes. This tables shows that from a gender perspective, men are more prone to cervical spine fractures. From an age perspective, the average of positive cases is 9 years greater than negative cases which suggests this injury is more common among the elderly~\cite{robinson2017c2}.

\subsection{Training Setup}
Training setup of the BLSTM~\cite{salehinejad2017recent} model was as follows: Adam optimizer~\cite{kingma2014adam} with a learning rate of $10^{-6}$; 100 training epochs; Batch size of 4; Input size of 2,048. Setup of the ResNet-50~\cite{he2016deep} model was as follows: Adam optimizer~\cite{kingma2014adam} with a learning rate of $10^{-3}$ and step decay of 5 with gamma 0.2; Number of input channels was 3; Random horizontal flip and rotation (10 degrees) augmentation method; Batch size of 16 and 50 training epochs. Early stopping and dropout regularization methods were applied. Given above setup, the proposed model was trained from scratch with randomly initialized weights and tested using a 7-fold cross-validation scheme.

\subsection{Performance Analysis}

Figure~\ref{fig:heatmap_cspine} shows two samples of axial cervical spine images with corresponding ground-truth masks and heatmaps generated using Grad-CAM~\cite{selvaraju2017grad} from the last layers of ResNet-50. The heatmaps show ResNet-50 can capture most  fracture areas with a relatively high false positive rate. 
Classification performance of the ResNet-50 at image level fracture detection in cervical spine axial images is presented in Table~\ref{T:cspine_resnet50}. In this experiment, cervical spine fracture is predicted just based on the spatial features in each axial image without counting temporal information. The results show $80.01\%$ and $77.61\%$ classification accuracy for the imbalanced and balanced datasets, respectively. However, the main drawback of this approach is the high level of false positives which leads to an inaccurate prediction for the entire case. Such that even if a single image in a case is false positive, the entire case will be false positive and vice-versa. These results show the importance of incorporating temporal features in training and inference.
Table~\ref{T:cspine_resnet50lstm} shows performance results of the combined BLSTM and ResNet-50 model at the case level for different number of LSTM units. The results regarding the imbalanced dataset show as the number of units increases from $96$ to $256$, the classification accuracy also increases from $75.79\%$ to $79.18\%$. However, the accuracy is approximately $71\%$ for the balanced dataset and it is less dependent on the number of LSTM units. 

Cervical spine imaging cases from a fracture detection perspective are highly biased from two perspectives. First is the larger number of negative cases when compared to positive cases. The second is generally a greater number of negative images than positive images within a positive case. These results clearly show the importance of considering these biases. For the imbalanced dataset, which reflects the natural distribution of negative and positive cases, as more LSTM units are utilized the imbalanced accuracy also increases. However, we observed no significant change in balanced accuracy. The higher performance on the imbalanced dataset is mainly due to the bias of the dataset toward negative cases and images and bias of the BLSTM layer in capturing the dependency between negative cases.

\section{Conclusions}
\label{sec:conclusion}
Automated fracture detection in cervical spine CT images is a very challenging task. In this paper, we proposed a machine learning model based on ResNet-50+BLSTM layer, which demonstrates capability of deep neural networks to address this challenge. We encourage the research community to work actively on this problem and we are in the process of releasing our large labeled dataset for research purposes.

\section{Compliance with Ethical Standards}
This study was approved by our Institutional Review Board with a waiver for informed consent. 

\section{Acknowledgments}
No funding was received for conducting this study. The authors have no relevant financial or non-financial interests to disclose. Authors have no conflict of interest.
\bibliographystyle{IEEEbib}
\bibliography{strings,mybibfile}

\end{document}